# SUSTAINED CARTOGRAPHIC INNOVATIONS IN NASCENT FRENCH CANADA

**Richard de Grijs**
*Department of Physics and Astronomy, Macquarie University,*
*Balaclava Road, Sydney, NSW 2109, Australia*
Email: richard.de-grijs@mq.edu.au

**Abstract:** Jean Deshayes, a teacher of mathematics in his native France, single-handedly put Québec on the map, literally. An accomplished astronomer, he used the lunar eclipse of 10–11 December 1685 to determine the settlement's longitude to unprecedented (although most likely fortuitously high) accuracy for the times. Deshayes contributed invaluable practical insights to the most important contemporary scientific debate—the discussion regarding the shape and size of the Earth—which still resonate today. Over the course of several decades and equipped with an increasingly sophisticated suite of surveyor's instruments, his careful scientific approach to hydrography and cartography of Canada's Saint Lawrence River is an excellent example of the *zeitgeist* associated with the 17th century's 'Scientific Revolution.'

## 1 QUÉBEC AND THE SAINT LAWRENCE RIVER

August 1685 marked the first arrival of Jean Deshayes in Québec, then a small but increasingly important settlement in New France.[1] Probably around 35 years old (*The Quarto*, 1974), Deshayes was bestowed the title of Royal Hydrographer[2] for the duration of a French fact-finding mission to the new colonies.[3]

Although his achievements in charting of the Saint Lawrence River have been discussed previously (e.g., Pritchard, 1979), here my aims are threefold: (i) I will explore his skill in using astronomical observations to determine accurate latitudes and longitudes. Deshayes' most famous achievement is his determination of the longitude of Québec to unprecedented accuracy for the times (although likely fortuitously high) based on lunar-eclipse timings.[4] He also obtained increasingly precise latitude determinations of a range of locales along the Saint Lawrence River using Polaris, the North Star, as celestial reference. Indeed, as we will see, Deshayes was sent to New France with the specific task to integrate astronomical observations for mapping purposes. (ii) To highlight Deshayes' evolving approach to hydrographic mapping, we will encounter a progressive line in the sophistication of the instruments employed, from his use of a simple quadrant during the November 1685 expedition, through use of that same quadrant combined with other surveyor's instruments in December 1685, to his application of a surveyor's plane-table equipped with telescopes in mid–late 1686. By the time of his death in 1706, he had acquired a suite of additional instruments, which had allowed him to make further corrections to his maps and charts, and to those of his contemporaries. (iii) Throughout this article, the reader will gain an understanding of Deshayes' role in charting the Saint Lawrence River and its shoals, and of the lasting legacy of his work. His life's achievement, a highly accurate chart of the Saint Lawrence River Valley, was reprinted a number of times throughout the 18th century (see Section 6). This chart subsequently formed the basis of numerous hydrographic, commercial and military maps. Addressing these aspects in tandem allows us to place his seminal contributions in the context of the times.

By the end of the 17th century, Jean Talon (1626–1694), Count d'Orsainville—the Intendant of Justice, Public Order and Finances in Canada, Acadia and Newfoundland (1665–1681)—was keenly aware of the need to carefully map the region's geography for the colony's fledgling economy to thrive. The Saint Lawrence River was to play a particularly important role in the colony's economic development.[5] However, at the time of Deshayes' commission there were no good hydrographic maps or topographical charts of the river's course, its banks and its surroundings. Given the Saint Lawrence River's often treacherous waters, the colony's need for competent pilots with access to reliable charts was urgent. The region's 1666 census made painfully clear that there were only 22 qualified *marins* (sailors) available for surveying and navigation in the entire territory (Archibald et al., 2009). Deshayes was dispatched to Acadia and Canada by King Louis XIV's right-hand adviser, the newly appointed Minister of the Marine, Jean-Baptiste Colbert (1619–1683). The scholar was given specific instructions[6] to obtain astronomical observations[7] and construct an accurate chart of the Saint Lawrence River.

Despite the mission's urgency, however, his reception in Québec was anything but welcoming. The acting Governor-General, Jacques-René Brisay (1637–1710), Marquis de Denonville, was clearly more interested in protecting the positions of Jean-Baptiste-Louis Franquelin (1650–c. 1712) as 'Royal Hydrographer in Québec' and that of his successor, Louis Jolliet (1645–c. 1700). This is abundantly clear from a passage in a letter to Colbert of 13 November 1685 (Gagnon, 1900–1901):

> If I had dared, I would have sent back to you the said Mr. des Hayes, because I believe that Mr. Jol[l]iet would have done this work well for you; but since I understand that he is here at your orders, it is not my place to question your reasons. Please have the kindness, my Lord, to let me know your orders, and whether you want to use Jol[l]iet [instead], who has his own boat.[8]

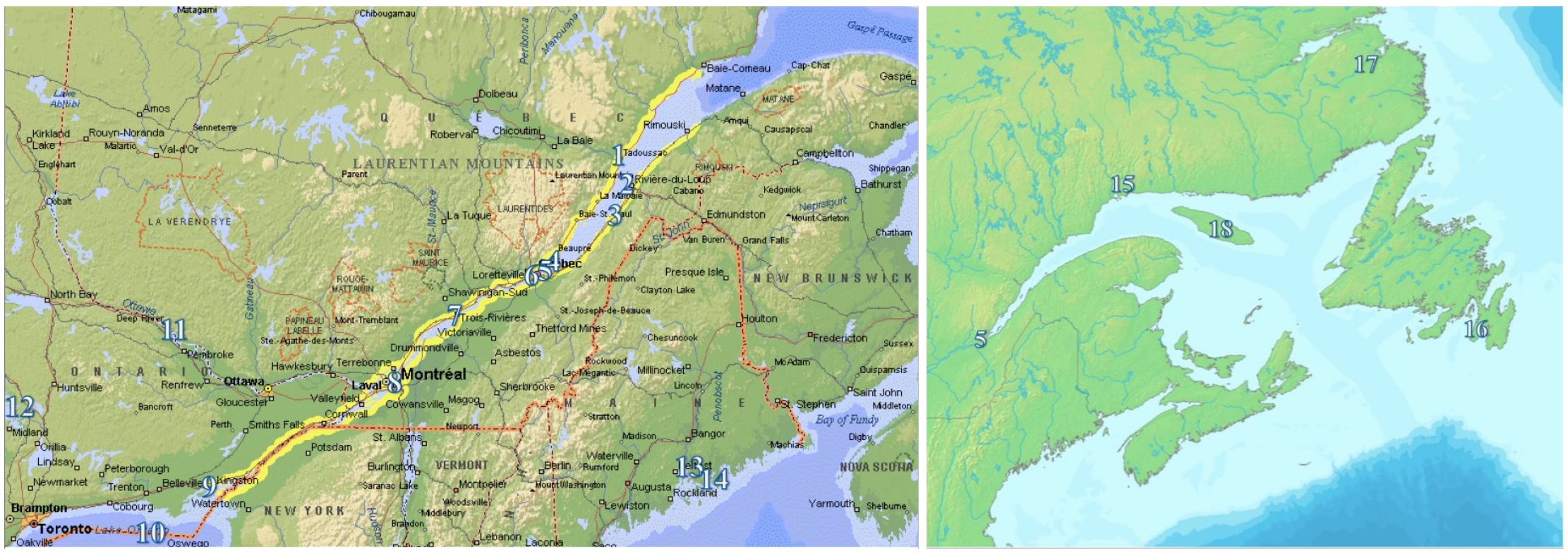

**Figure 1**: Locations of relevance to Deshayes' narrative. 1: Tadoussac (Montagnet); 2: Île aux Lièvres; 3: Rivière Ouelle; 4: Cap Tourmente; 5: Île d'Orléans; 6: Québec (including Pointe-Lévy); 7: Trois-Rivières; 8: Montreal; 9: Fort Frontenac (Cataracouy); 10: Lake Ontario; 11: Ottawa River; 12: Georgian Bay; 13: Pentagoûët; 14: Acadia; 15: Sept-Îles; 16: Plaisance (Newfoundland); 17: Labrador; 18: Île d'Anticosti.

Despite his apparently hurt feelings, shortly after Deshayes' arrival, Denonville took the cartographer on a physically challenging 350-mile canoe ride up the Saint Lawrence River to its source at Fort Frontenac, Cataracouy—present-day Kingston, ON—on Lake Ontario (see Figure 1). The Frenchman's main task was to derive the latitudes of all places where he went ashore, a key task for all explorers of New France since the arrival in the colonies of Samuel de Champlain (Jarrell 1988). The latter had based his (rather uncertain[9]) latitude determinations on solar altitude measurements with his astrolabe, whereas Deshayes relied on measuring the altitude of Polaris above the northern horizon while accounting for the small daily circle Polaris traces around the celestial North Pole owing to the Earth's rotation. The enterprise turned sour soon after their departure, however, as we learn from Denonville's report (Delanglez, 1943):

> This was not done without trouble. I thought that he was going to die at any moment in my canoe. I told him to mark on his map all the inhabited places of the colony. You may be sure, my Lord, that it is accurate. In time, my Lord, we shall try to provide you with a good knowledge of the whole country.

In fact, the expedition was pressed for time because of Denonville's desire to reach Lake Ontario as soon as possible. Deshayes was therefore not in a position to properly establish base stations for his angle measurements (Pritchard, 1979). Using a quadrant he was only able to determine the approximate latitudes of ten places where he managed to go ashore. This was a crude measure by any means, even for the times.[10] Limited by both the instrument's relatively small size (and, hence, by the angular resolution attainable) and external influences (including wind and wave motions), it allowed him to only determine positions to within about five minutes of arc. At Fort Frontenac, he obtained just three measurements owing to high winds. In modern terms, Deshayes' latitudes were systematically too far north by between 5' and 25' (Pritchard, 1979). This can be attributed to either observer error—systematic positional offsets between the quadrant's viewing angle and the observer's eye line, elevation differences or other external influences—or instrumentation. It is unknown what caused the main offsets in this case, although one of Deshayes' successors attributed the systematic offsets to the instruments that had been employed.

## 2 THE DECEMBER 1685 LUNAR ECLIPSE

In a letter of 3 December that year, Denonville re-emphasized Deshayes' frail health to Colbert: "Mr. Deshayes, whom you sent to compile the river's map is so exhausted by illness that he will not be able to work all year" (Roy, 1916: 129). However, just a week later Deshayes made one of his most important contributions to the development of cartography in New France.

On 10 December 1685, he carefully observed the time when the Moon left the shadow of the Earth during a full lunar eclipse,[11, 12] which—according to modern calculations[13]—commenced at 23:22 UT. Back in Paris, Jean-Dominique Cassini (1625–1712), the founding director of the Observatoire de Paris, observed the same eclipse (Chartrand et al., 1987). Their time measurements[14] differed by 4 hours 48 minutes and 52 seconds, corresponding to a distance in longitude of 72°13' between Paris and Québec. These values were published in the *Connoissance des temps pour l'année 1706 au méridien de Paris*, the leading almanac in the francophone world.[15]

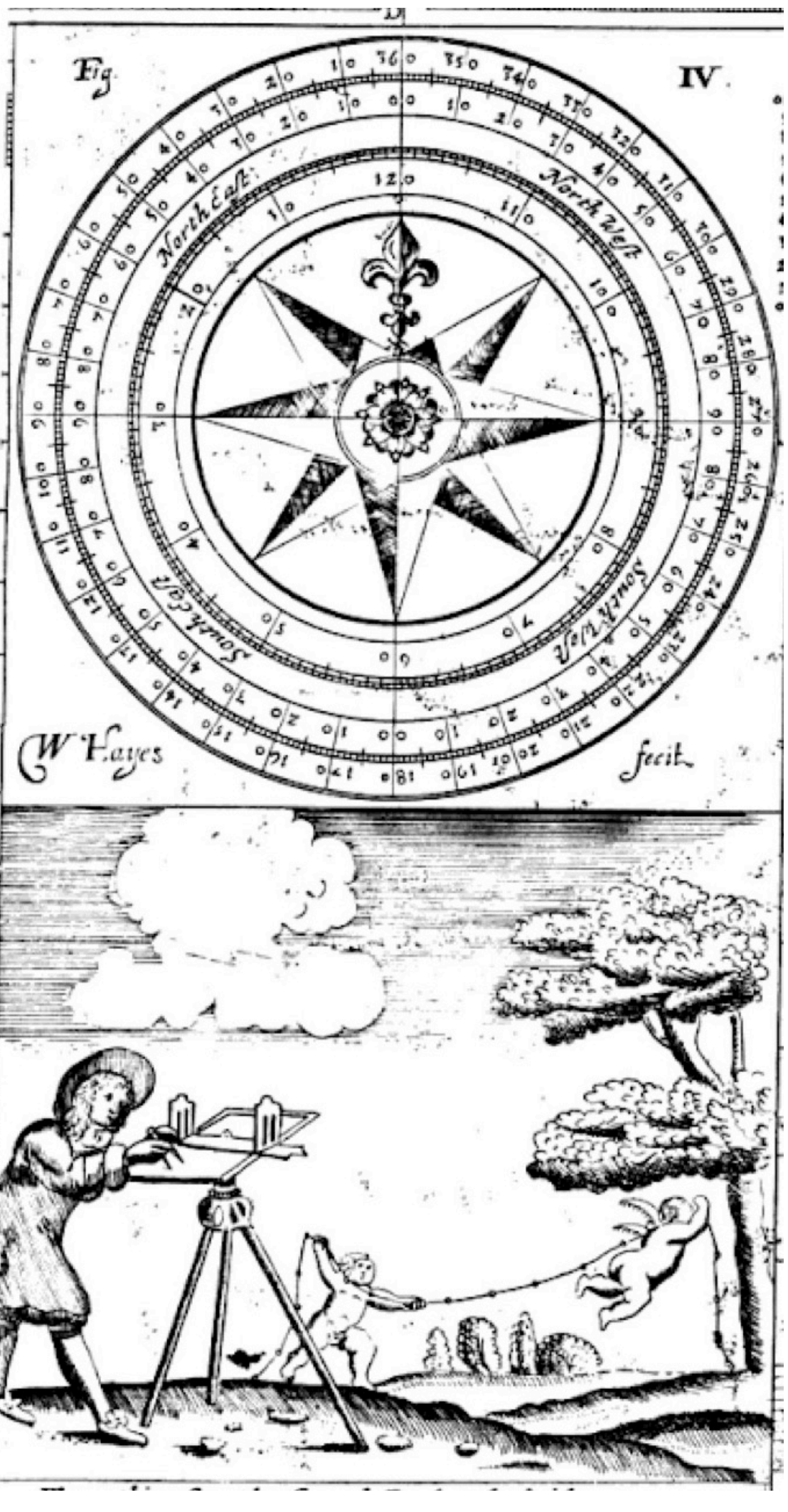

**Figure 2**: Surveyor's simple plane-table similar to that initially used by Deshayes (bottom), combined with (*top*) a sketch of its operational surface ('Fig. IV'). (Leybourn, 1722: LXXXIX).

There is some disagreement among scholars as to whether Deshayes determined the longitude of Québec himself (Pritchard, 2004; Palomino, 2009) or whether this was instead achieved by Cassini at a later date. Both men likely performed the calculations independently: on a copy of a document accompanying his new chart of the river, we find handwritten notes by Deshayes himself, explaining how he determined the settlement's longitude.[16] The 1685 lunar-eclipse-based meridian determination is often thought to have been the first *accurate* longitude determination in New France. However, Pritchard (1979) suggests that Deshayes' *subsequent* determination of the Québec meridian in the spring of 1686 (see below) should be regarded the first such measurement. The latter appears to have been published and widely advertised well before the lunar-eclipse-based determination reached a wider audience, although it may not have been as accurate.

Assisted by the Jesuit Father Pierre Raffiex (1633–1724), Deshayes observed several positional angles from the belfry of the Jesuits' chapel toward unspecified "notable places." He subsequently observed the height of the Sun, as well as its rising and setting times, from some of those locations on two consecutive days. Deshayes obtained his measurements using simple and box compasses, a circle, a quadrant and a surveyor's 'plane-table' (see Figure 2). In his seminal work, *The Compleat Surveyor*, the English mathematician and surveyor William Leybourn (1626–1716) specifically recommends that "the manner of Surveying by way of Traverse"—establishing control networks of survey stations which each serve as calibration base for the next station's position determination—is best suited to map large areas, including entire countries. In this case, "the Account were best to be kept by the bearing and distance of Places; which way of Surveying is shewed at large in the *Fourth Book*": see Figure 2 (top), where the "Card" placed on the surveyor's plane-table is

> … divided into four quarters, divided each into 90 Degrees, and numbred [sic] from the North and South, towards the East and West, by 10, 20, 30, etc. to 90 Degrees, 00 Degrees standing at the North and South points, and 90 Degrees at the East and West points, and then will the four quarters of the Card be thus denominated, the North-East, North-West, South-East and South-West Quarters.

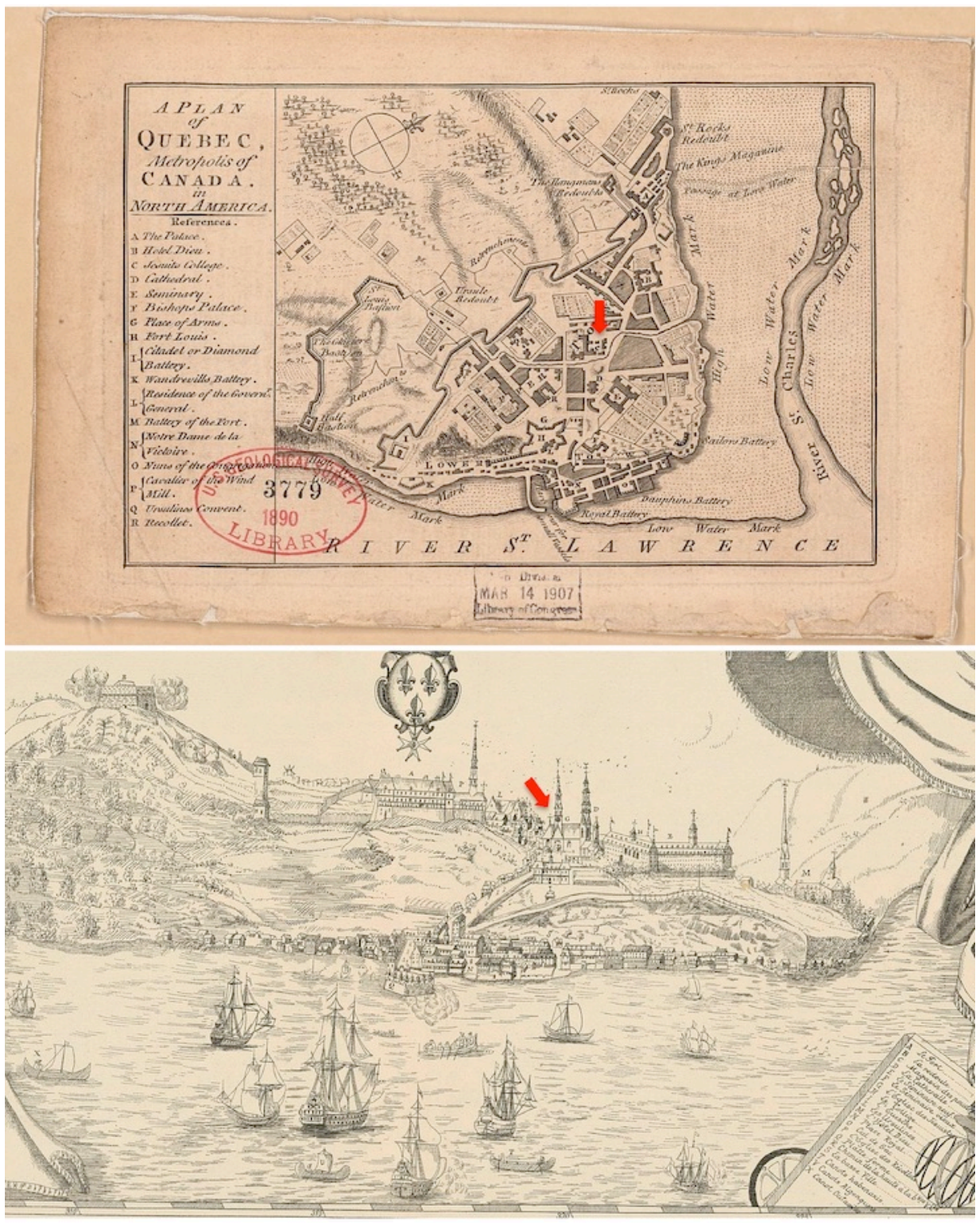

**Figure 3**: Location and profile view of the Jesuit College in 17th and 18th century Québec City (red arrows): (*top*) *"A plan of Quebec, metropolis of Canada in North America"* (*London Magazine*, April 1759; courtesy Geography & Map Division, US Library of Congress, digital ID g3454q.ar060000; public domain). (*bottom*) *"Québec vue de l'Est,"* detail (Charles Bécart de Fonville, 1699; courtesy Bibliothèque Nationale de France, public domain).

Numerous and varied measurements obtained with Deshayes' basic instruments, combined with the latitude of Québec determined by Father Raffiex in the garden of the Jesuit's quarter on an earlier occasion, allowed the men to establish a meridian for Québec. The longitude meridian was taken from Dupont's windmill, located at Cape Diamond, through the chapel's belfry, to a mountain peak some 12 to 15 leagues to the north. Deshayes emphasized that "[t]he observed azimuth is the foundation of the whole chart" (Pritchard, 1979: 132). Nevertheless, the accuracy of the earlier, lunar-eclipse-based meridian determination of Québec, reported to a precision of seconds of arc, was far superior. Lunar-eclipse timings (particularly ingress and egress of the Earth's shadow) could be determined with high accuracy using contemporary clocks. Pritchard (1979) asserts that the actual distance between Deshayes' longitude determination and the true meridian is only 1°23′. This is equivalent to 102 km at the latitude of Québec;[17] the discrepancy may have been caused by both observational errors and deficiencies in the lunar ephemeris tables at Deshayes' disposal (Jarrell, 1988).

It is unclear exactly where in Québec City Deshayes obtained his observations. If he used the Jesuit chapel's belfry as his base, present-day cartography renders its position at 71°12′29″ East (the Jesuit Collège was located at the city's present-day City Hall[18]; see Figure 3). Combined with the Paris meridian through the Observatoire de Paris of 2°20′14.03″ West, this yields a geographic separation of 73°32′43″, compared with Deshayes' determination of 72°13′, a difference of 1°20′. Alternatively, if we use the modern longitudes of Québec City and Paris, respectively 71°13′ West[19] and 2°21′ East[20], we find a longitude difference of 73°34′, which deviates by 1°13′ from Deshayes' value. It thus transpires that an uncertainty of order 7′ in contemporary measurements, corresponding to approximately 9 km at the latitude of Québec City, may not have been all that unusual. This suggests that Deshayes' declaration of the accuracy of his longitude measurement to within a few minutes of arc, and particularly with respect to the longitude of Paris, may have been overly optimistic.

Deshayes' highly accurate establishment of the Québec meridian is not just an interesting historical side-note. French efforts to determine the shape of the Earth were the focus of a fierce scientific debate raging in the *Académie Royale des Sciences* in the 16th to 18th centuries. Political, commercial and military considerations were only secondary to the main scientific problem at hand (Pritchard, 2004). We will shortly return to the scientific implications of these developments. Nevertheless, and despite this emphasis on resolving one of the key scientific problems of the era, Deshayes' 1685 expedition to New France ultimately had a practical aim, that is, to construct a hydrographic chart of the Saint Lawrence River. It is therefore highly significant that the mission was sponsored by the King's navy rather than by the *Académie*. This is particularly so, because the institution had been established by Colbert with the ultimate aims to give the French government, and in particular its King, more prestige and authority among the European nations and to make France a great maritime power. In essence, therefore, the *Académie* responded to the needs of the State as regards technical policy advice. Its scientists engaged in both full-time scientific research and government-mandated engineering projects (Gauja, 1949; Saunders, 1984). As such, the *Académie*'s scientific efforts were not truly independent, but they relied to a large extent on the political desires of the ruling monarch and his ministers. Although Colbert supported scientific exploration as a goal in itself, he was particularly keen to pursue those aspects of mathematics, astronomy and mechanics that would benefit the French navy's navigation capabilities (Gauja, 1949; Saunders, 1984). In turn, this would strengthen the navy and expand the State's commercial reach around the world.

## 3 THE RIGHT MAN FOR THE JOB

Land measurements are important to define ownership rights. The precise meridian based on Deshayes' measurements would serve for cadastral mapping (Heidenreich, 2006), that is, to design surveys for land grants, known as *seigneuries* in the colonies. The first regional map indicating property boundaries and building locations was the 1641 map of the surveyor Jean Bourdon (c. 1601–1668). As we will see below, the first reliable chart of the Saint Lawrence River Valley was constructed by Jolliet and Franquelin in 1685. It was subsequently improved by Deshayes.[21] Deshayes was well placed to lead the colony's cartographic mapping efforts. And despite his initial misgivings, Denonville was apparently impressed by Deshayes' map-making and charting skills, on which he expounded in his letter to Colbert of 8 May 1686 (Gagnon, 1900–1901). In the same letter, he also requested resumption of hydrography courses in Québec:

> I have the honor to tell you that it would be very useful for the King and the country to have here someone teaching the principles of navigation. The Jesuit Fathers had one person who did a good job, who did everything we needed to educate people in navigation and who served our merchants [well]. It has been some time since this man died; unless the King gives us someone to replace him, we will not have anyone [competent] anymore; however we are short of pilots, five having died in the past three years, and the country suffers a lot.

Although nothing is known about his early life or family background, references to a certain Deshayes (Deshaies) first appear around 1668. If this 'Deshaies' was indeed the Royal Hydrographer of 1685, he was clearly already recognized as a promising scholar.[22] The engineer Sainte-Colombe described him as "a man of strong good sense" (Pritchard, 1969; revised 1982).

Mathematics education in 17th century France was still predominantly the realm of the Roman Catholic church. Essentially all French universities were controlled by teaching congregations, the best-known and most powerful of which during the first half of the 17th century—until it was eventually dissolved by King Louis XV in 1764 (Vogel, 2010)—was the Society of Jesus: by the middle of the century, the Society was in charge of educating some 14,000 pupils in Paris alone, whereas it ran 612 Jesuit colleges across the country (Sofroniou, 2016). In 16th and 17th century France, a more practical branch of mathematics pedagogy developed, aimed at middle-class students, with particular emphasis on the vocational use of mathematics in relation to commerce, surveying and navigation (Jones, 1967). As an example, these new developments are clearly evidenced in Father Petrus Galtrucius Aurelianensis' (1602–1681) *Mathematicae Totius* (Cambridge, 1683), which included sections focusing on the four 'quadrivium' subjects, arithmetic, geometry, music and astronomy, as well as four practical subjects, chronology, 'gnomonicae' (sun dials), geography and optics. Deshayes was educated and taught mathematics and hydrology in this broader national context. Deshayes' grounding in the Jesuit teachings appears to be further supported by his evidently strong links, post-1685, to the Jesuit Collège in Québec, which represented the main seat of knowledge and education in the young colony of New France.

In the late 1660s, the *Académie*, and Colbert in particular, was keen to test a novel method for longitude determination at sea acquired from a foreign scholar—possibly the Dutch natural philosopher Christiaan Huygens (1629–1695; de Grijs, 2017). Deshayes was tasked with validating the method—which involved obtaining time measurements of eclipses of the Moon and of Jupiter's satellites—on a voyage to Acadia, set to take place shortly after 1668. Given the immature status of timekeeping onboard rolling and pitching ships at the time, it is unclear whether Deshayes achieved his aims. Nevertheless, in 1669 he presented his own, rather complicated approach to determining one's longitude at sea based on lunar-distance measurements. It was never widely adopted, however, because of the less-than-straightforward calculations sailors would need to perform to obtain useful results (Pritchard, 1979).

Meanwhile, Colbert received a letter from his protégé Jean Talon, announcing that (Olmsted, 1960; see also O'Connor and Robertson, 2012)

> I have already had the honor to write you that [Jean] Richer will embark … on the *Saint Sébastien* for Acadia. Deshayes will also sail on the same vessel with the instrument[23] that he has made in Paris. It is to be hoped that from the contact of these two men, who are embarking on good terms, knowledge will result with which you may be satisfied.

Delayed by inclement weather conditions, the *Saint Sébastien* eventually sailed from La Rochelle, France, on 1 May 1670. Deshayes is not mentioned again in the expedition's reports. However, the scientific aspects of the voyage were not altogether successful (de Grijs, 2017), so that Deshayes may simply not have obtained any worthwhile longitude determinations. Nevertheless, the expedition resulted in a number of important cartographic measurements, particularly that of the *latitude* of the French fort at Pentagoûët (present-day Castine, Maine) of 44°23'20" North, only 5" from the correct value, 44°23'25" North. This measurement is significant and the accuracy remarkable, because it was given to the nearest second of arc at a time when such measurements were usually quoted in degrees and minutes only.

Deshayes' efforts in determining accurate positions were part of a larger French scheme driven by Cassini and the *Académie*, once again combining scientific exploration with the State's practical needs. King Louis XIV had appointed teams of astronomers to map his kingdom using observations of stellar occultations or eclipses (de Grijs, 2017). Jean Picard (1620–1682), founding member of the *Académie*, and Philippe de la Hire (1640–1718) determined the coordinates of the French Atlantic coast between 1679 and 1681, which led to significantly revised maps. Meanwhile, Deshayes and someone known as 'le Sieur Varin,' possibly Charles-Nicolas Varin (Danson, 2006), a mathematics teacher from Paris, contributed observations in Rouen and Dieppe, before setting off on a voyage to Africa and the West Indies in 1681–1683. Deshayes was promoted to His Majesty's Engineer for Hydrography and dispatched to determine longitudes on an extended voyage to the island of Gorée (off the African coast near present-day Dakar, Senegal) and to Martinique and Guadeloupe in the Caribbean.

This latter expedition was of great importance, since King Louis XIII had already issued a decree on 1 July 1634 announcing that French scientists and navigators should use the meridian through the westernmost of the Canary Islands, Ferro (El Hierro), as positional reference. Unfortunately, at the time the longitude of the Canary Islands with respect to that of Paris was poorly known (Dew, 2010). The expedition of Deshayes and Varin was meant to obtain a first firm measurement of Ferro's longitude. However, after significant delays, the idea of sailing to Ferro was abandoned and the expedition's focus was recast to visiting

Gorée Island and the Antilles instead (Dew, 2010).

Despite this apparent setback, the expedition provided crucial measurements of the length of the simple 'seconds' pendulum, which beat with a precisely calibrated period and might be employed as a universal time standard. However, the Earth's rotation, combined with its slightly flattened shape at the Poles (a shape known as an 'oblate spheroid'), reduces the net gravitational force component at the Equator compared with that at the Poles. As a consequence, the length of the seconds pendulum should become shorter as one travels toward lower latitudes. Its length in Paris was well-established (de Grijs, 2017). A number of leading scholars associated with the *Académie*—including Huygens, Richer and Picard—had measured highly consistent values for the pendulum's one-second fall length in Paris (Harper, 2011: Table 5.2). Away from Paris, the available measurements were not as clear-cut, however. Richer's measurements in 1672 in Cayenne, near the Equator, implied a shortening by 1¼ *lignes* (2.8 mm). Confusingly, in 1682 the team led by Varin on the expedition to Gorée Island and Guadeloupe concluded that the seconds pendulum had to be shortened by 2 *lignes* on Gorée Island (Varin et al., 1682), 10° further north than Cayenne, while an additional correction was required in Guadeloupe, more than 16° north of the Equator (de Grijs, 2017). Given these inconsistencies, a systematic exploration of the length of the seconds pendulum and its variation with latitude was urgently needed, using the same experimental set-up. This realization emphasizes the changing scholarly attitudes driven by the 16th and 17th century 'Scientific Revolution', when a critical, empirical assessment of nature rather than religious doctrine or Aristotelian explanations became the *modus operandi*. Deshayes' contributions to measuring the length of the seconds pendulum in both Gorée and the West Indies thus provided crucial information regarding the shape of the Earth.

In this context and given his credentials as a careful scholar, Deshayes was clearly well positioned to take the lead in charting the Saint Lawrence River Valley based on careful astronomical observations. Despite his initial health woes, preparations proceeded decisively during the winter of 1685–1686. He undertook extensive consultations with local navigators, including Jolliet, in an effort to understand the best approach to surveying the land. He established the baselines for his large-scale triangulation survey of the river the following year, with his first baseline running a distance of 1560 *toises*[24] from "la pointe des roches" in Lower Town Québec across the frozen river to the second knoll of Pointe-Lévy (Pointe-Lévis). He established two additional baselines, one of 608 *toises* and a second of 6805 *toises* (close to 8¼ miles, or just over 13 km). The former was laid from the same point of origin as his first baseline, to the corresponding shoreline on the Québec side, while the latter extended downriver along the north shore opposite the Île aux Lièvres. Eventually, his survey of the Saint Lawrence River Valley covered the entire region from Québec City downriver to the area known today as Sept-Îles. It required measurements of some 300 triangles spanning a distance of around 560 km (Hamilton and Seibert, 1996).

Perhaps surprisingly, Deshayes' three baselines were the least accurately measured distances of the entire survey—possibly on account of his rudimentary measurement device: all he had access to was a *demi-pied de Roi* (a six-inch rule), which in first instance he applied to construct a *toise*, in turn used to measure a 10-*toise* or approximately 19.49 meter-long cable covered in beeswax (to reduce the effects of variations in humidity; Pritchard, 1979). Nevertheless, this device would have rendered determining any significant distances cumbersome even in optimal conditions, let alone those across the frozen Saint Lawrence River. Although one would ordinarily expect that any subsequent measures would be even less accurate, Deshayes managed to correct for most of the inaccuracies that would inevitably propagate into his work by employing independently observed latitudes as calibration benchmarks. For a detailed account of how Deshayes obtained his measurements, as well as of the difficulties thrown up by the rugged terrain, I refer the reader to Pritchard (1979), where we learn that …

> [o]bserved angles were extremely important but difficult to establish in the uninhabited region of the lower St. Lawrence where few convenient objects existed on which to take bearings. The great width of the river made accurate definition of objects on the opposite shore impossible and Deshayes had to sight on distant mountain summits in order to obtain guides to further stations. The possible error was very great, for although the hydrographer established stations and turned angles on all the headlands along the north shore, the mountains to the south were necessary to tie the stations together. The summits were irregular and the same point on any summit was not observed from any two stations. (…) Much of the shoreline had to be located by compass bearings which introduced additional errors owing to magnetic declination, but Deshayes tried to correct these by incorporating spherical observations.

With the expedition about to start, on 8 May 1686 Denonville wrote to Colbert (Roy, 1916),

> Mr Deshayes is waiting for the caulking of his boat[25] to be completed in order to travel down the river and survey the region as far as the Gulf of Saint Lawrence; in a few days he will set sail. (…) Mr Deshayes has gone for a hike along the banks of the river below Québec, while waiting for the navigation channel to clear for him to sail down to the river's estuary: he will focus on identifying difficult passages and places where islands would prevent ships from coming here in case we had to fear a powerful enemy approaching from the sea.[26]

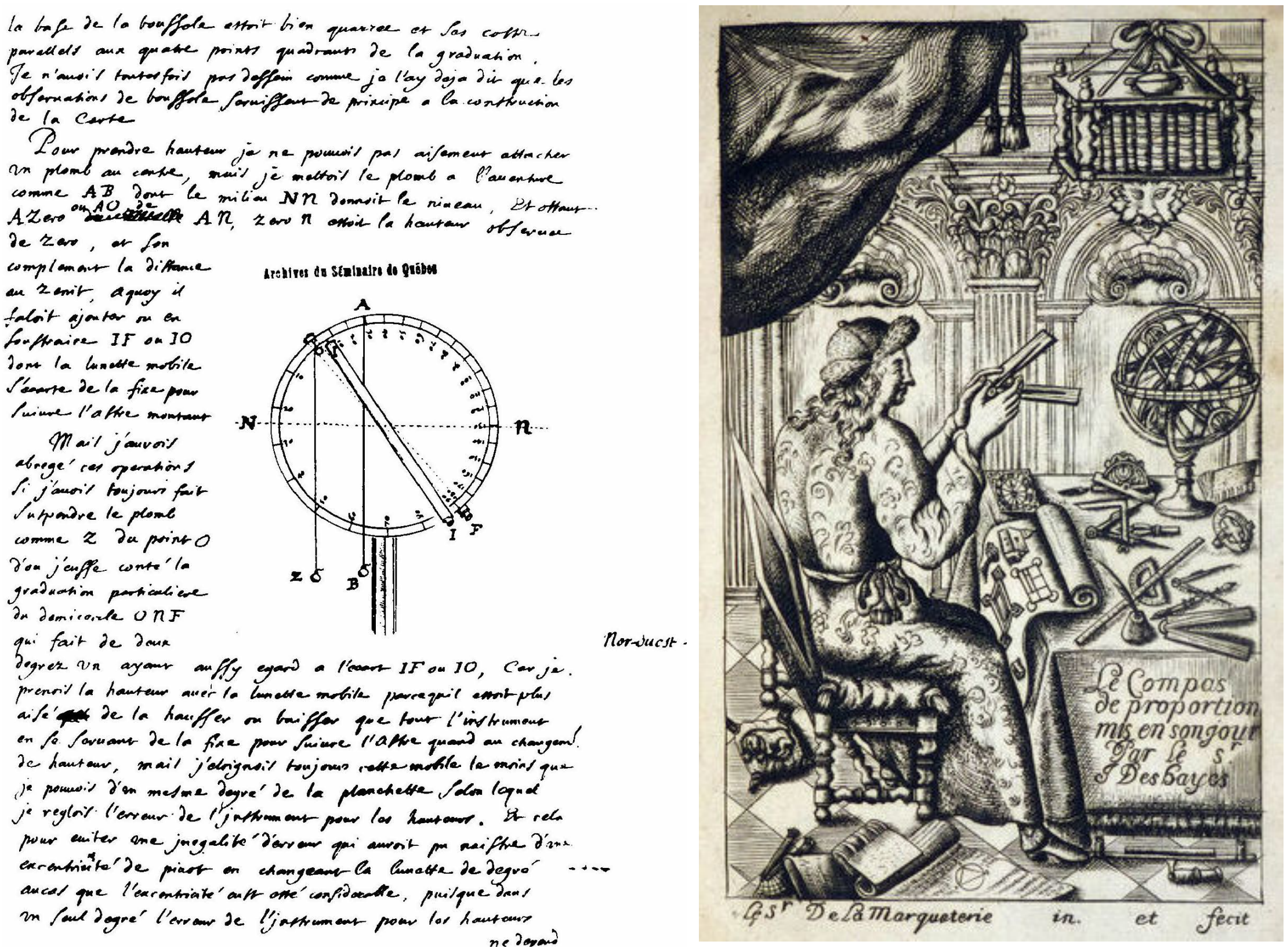
la base de la boussole estoit bien quarree et ses costes paralleles aux quatre points quadrants de la graduation.
Je n'avois toutesfois pas dessein comme je l'ay deja dit que les observations de boussole servissent de principe a la construction de la Carte
Pour prendre hauteur je ne pouvois pas aisement attacher un plomb au centre, mais je mettois le plomb a l'aventure comme AB dont le milieu NN donnoit le niveau, Et ostant A Zero ou AO de AN, zero N estoit la hauteur observee de Zero, et son complement la distance au Zenit, a quoy il faloit ajouter ou en soustraire IF ou IO dont la lunette mobile s'ecarte de la fixe pour suivre l'astre montant
Mais j'aurois abrege ces operations si j'avois toujours fait suspendre le plomb comme Z du point O d'ou j'eusse conte la graduation particuliere du demicercle ONF qui fait de deux degrez un ayant aussy egard a l'ecart IF ou IO, Car je prenois la hauteur avec la lunette mobile parcequ'il estoit plus aisé de la hausser ou baisser que tout l'instrument en se servant de la fixe pour suivre l'astre quand au changement de hauteur, mais j'eloignois toujours cette mobile le moins que je pouvois d'un mesme degré de la planchette selon lequel je reglois l'erreur de l'instrument pour les hauteurs. Et cela pour eviter une inegalité d'erreur qui auroit pu naistre d'une excentricité de pivot en changeant la lunette de degré au cas que l'excentricité eust esté considerable, puisque dans un seul degré l'erreur de l'instrument pour les hauteurs ne devoit

**Figure 4**: (*left*) Deshayes' description of his plane-table instrument (1686). (Courtesy of the Archives du séminaire de Québec, Université Laval, *Polygraphie* 2, no. 34, f. 18; not in copyright.) (*right*) Deshayes surrounded by his tools of trade. (Henrion and Deshayes, 1681; public domain.)

Clearly, Deshayes had recovered his health by this time, given that he made extensive journeys across the frozen landscape on snowshoes. He hiked as far south as the Rivière Ouelle, some 110 km downstream of Québec City, while his northern pursuits reached Cap Tourmente and the Île d'Orléans (Pritchard, 1979). Deshayes spent the next six months, until November 1686, surveying the river below Québec, accompanied by a crew of seven. Using up-to-date instruments—including a plane-table equipped with fixed and moveable sighting telescopic alidades (sighting devices for determining angles), a major innovation at the time—and surveying techniques, he compiled a detailed hydrographic map of the river valley, the Saint Lawrence Seaway, including remarkably accurate river soundings (see Figure 4).[27]

> I had a plane table made by But[t]erfield[28] only nine to ten inches in diameter on which was attached a brass arm divided into degrees.[29] There were two telescopes for sighting, passing through the center [axis], the one mobile on the graduated side, the other fixed to the other side. (Deshayes, 1686)

Pressed for time, however, he followed a surveying regime that was “more forced than natural and deliberate” (Pritchard, 1979: 128; transl.) It consisted of alternating sounding measurements of the river with trigonometric readings on shore, to varying degrees of detail—depending on the importance of the specific location. He also obtained 21 latitude measurements below Québec, as before based on sightings of Polaris. Again, his values were systematically too large, but this time only by one to five minutes of arc. It is most likely that his instrument was to blame for this systematic offset. At least, that was the conclusion reached some 70 years later by Father Joseph Pierre de Bonnécamps (1707/8–1790), the colony’s last Professor of Hydrography. However, de Bonnécamps added (Gosselin, 1897: 98; transl.),

> … if you had, as I do, the instrument that he used, your surprize [sic] would soon cease. It is a plane-table 8-2/3 inches in diameter and fitted with a copper edge divided into 360 degrees each of which is at least 5/6 [of a degree] out of line.[30] Now, however experienced an observer might be, could he be responsible for 7' or 8' with such an instrument?

## 4 AN INCOMPLETE CHART AS SURPRISING BENCHMARK

For unknown reasons, Deshayes left for France in November 1686, without having completed his survey. The chart resulting from Deshayes’ expedition did not include the Île d’Anticosti or the Île Percé. Denonville informed the minister of the survey’s status in a letter dated 10 November 1686 (Roy, 1916: 130):

> Mr Deshayes will report to you, my Lord, on his job. We would need him to continue for several years in the Gulf [of Saint Lawrence] and at the estuary at the Grand Bank, given that we do not have any reliable maps; there is an infinity of currents to observe which change from side to side, with many incorrectly placed islands [located] on the maps.

On 16 November, Intendant de Meulles added (Roy, 191: 130),

> I have also paid to Mr Deshayes, hydrographer, 150 l. [livres] to cover the expenses of his voyage from France. If by doing so I have done something which is unpleasant to you, my Lord, I will not do it again.

Deshayes did not return until 1702. Meanwhile, Franquelin took on Deshayes’ role as Royal Hydrographer and became the colony’s first official mapmaker. His maps are known for their colorful and decorative features, but not for their scientific accuracy.

The importance of the hydrographic map resulting from Deshayes’ 1686 triangulation survey[31] cannot be overstated: as we will see below, it became the precursor of all modern Canadian charts. It was likely used by James Cook (1728–1779), the well-known British explorer, in 1759 on the Saint Lawrence River in Québec’s Lower Town (Cook’s later charts would have supplanted Deshayes’ efforts of the previous century, as the British took control of North America; see Section 6). Yet, at the time of its compilation, little was done with the information obtained by these early expeditions. It took until the end of the century before Deshayes’ practical insights reached those who needed them. The main problem was the ready and cheap availability of reliable Dutch charts for crossing the North Atlantic (although those lacked detail beyond coastal areas) and—perhaps surprisingly given the dual practical and scientific aims foreseen for the young *Académie*—the absence of a coordinated scientific approach within the French establishment prior to the end of the 17th century (see below).

In 1699, the *Académie* praised Deshayes’ chart, “which traces the Saint Lawrence River’s course, from its mouth to Lake Ontario [and] has been judged very accurate [and] of great use for navigation of the Saint Lawrence River.”[32] It covered the river’s estuary in detail, from the Pointe-des-Monts Peles at the onset of the Gulf of Saint Lawrence to the Coste de l’Île Percé on the Gulf’s opposite shore. A large inset covered the river valley from Québec to Lake Ontario. Contrary to common practice at the time, Deshayes’ chart did not include a decorative title cartouche. However, the cartographer had instead provided a detailed description in an accompanying document.[33] The idea was not only that this document could be used to understand the chart’s features, but also that it would be a teaching aid as well as a tool for pilots navigating the river. Deshayes’ report is composed of two parts, with the first part containing the survey summary and sailing directions, and the second part "the rest, [which] is only to render witness to the fidelity of the measurements of this chart." However, 'the rest' is what makes this chart and its documentation particularly valuable.

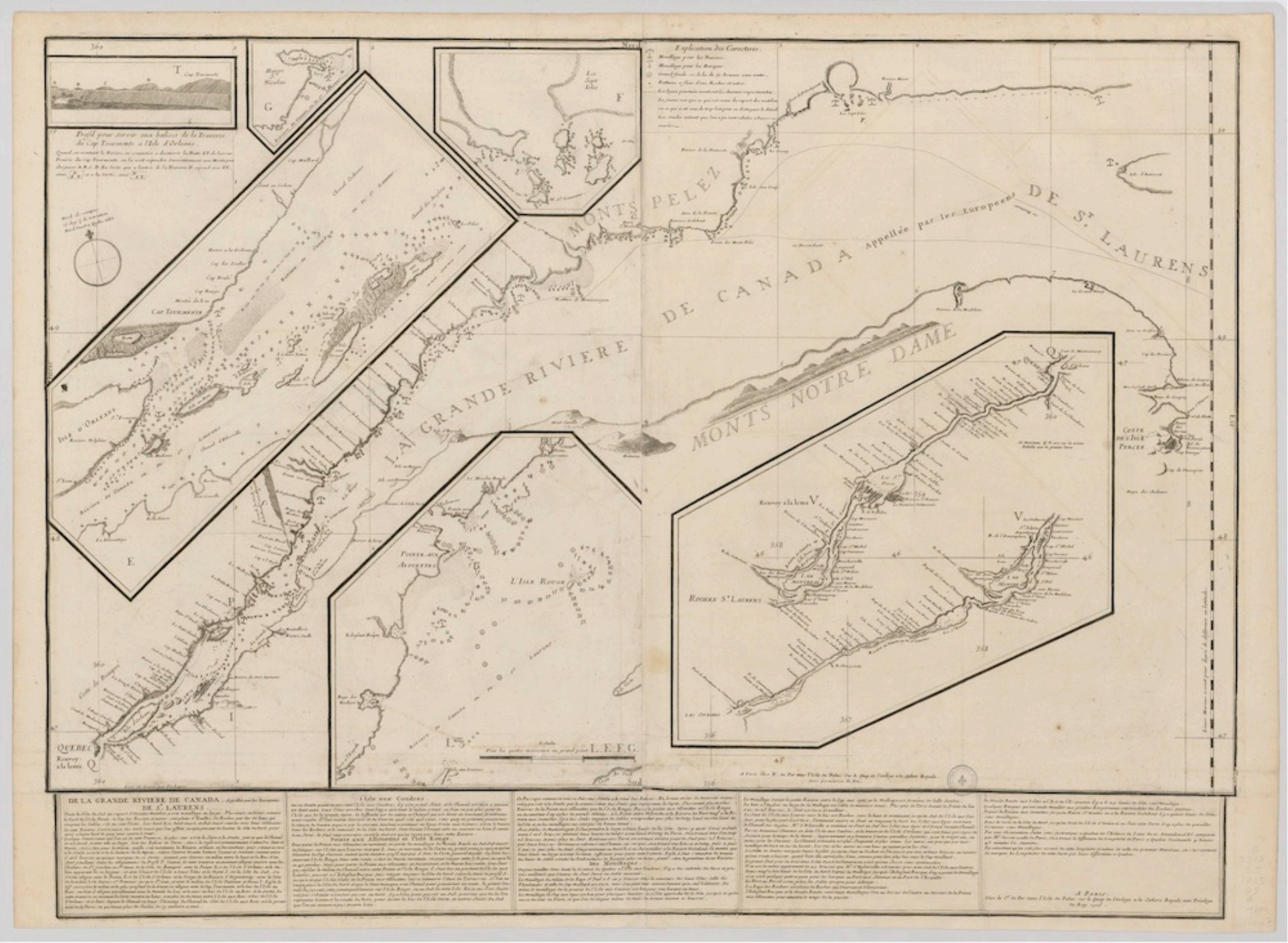

**Figure 5**: Deshayes' 1715 chart of the Saint Lawrence River Valley, *De la grande riviere de Canada appellée par les Europeens de St. Laurens* (Paris: Nicolas de Fer) *Credit: BAnQ Rosemont–La Petite-Patrie, G 3312 S5 1715 D4 CAR* (Permission to reproduce for research purposes; http://numerique.banq.qc.ca/apropos/conditions.html; for a very high resolution map showing intricate details, see http://cartanciennes.free.fr//maps/stlaurens.jpg).

Figure 5 represents the 1715 version of Deshayes' chart, showing the great level of detail. The chart is based on equal-angle projection and oriented with geographic North pointing upward. The direction to the magnetic North Pole is shown by means of a wind rose and the inscription "Nord de compas 15 deg. ½ de variation Nord-Ouest à Québec 1686," implying that the geographic and magnetic North Poles differed by 15½ degrees to the northwest on local maps. Deshayes adopted the Québec meridian he had determined based on the December 1685 lunar eclipse as his positional reference. He also included a longitude–latitude grid on a Mercator map projection, as had been in common use since the mid- to late-16th century (e.g., de Grijs, 2017), but not a linear scale. In 1703, the French cartographer Guillaume Delisle (1675–1726), known for his careful cartographic research, published the first map of the region to contain a modern grid pattern.

Deshayes' chart is rich in detail: it includes anchorages, deep seabed, reefs (flats, rocks and others), channels and soundings (reduced water levels at low tide). One can also discern houses and mills on the river banks. The chart's font size and form (regular, italic, straight and slanted) are used as tools to differentiate among types of landmarks: rivers, canals, rapids, coves, bays, headlands, capes, islets, islands, ports or harbors. Sounding depths and reefs are indicated throughout. Differences in elevation are illustrated by hatched lines, marshland areas are dotted with blades of grass and sandbanks are indicated by dotted areas. These latter were, of course, of great interest to the river's navigators; they include areas near the Île de Saint Barnabé and the Pointe-aux-Allumettes. As a practical navigational aid, four of the five insets focus on navigation directions in regions that are particularly difficult to navigate, including the Pointe-aux-Allumettes, the navigation channels surrounding the Île d'Orléans, the Sept-Îles area and the harbor of Saint-Nicolas. The fifth inset represents two sections of the river's course, from Québec City to Montréal and from Montréal to Lake Ontario. This latter inset does not contain information about potential navigation hazards.

Although it is commonly believed that Deshayes' chart was not published until 1702, an unofficial version may have appeared already in 1695 (*The Quarto*, 1974; Pritchard, 1979).[34] While local pilots and navigators worked hard to characterize navigation hazards, their charts were relatively crude compared with Deshayes' monumental effort. Jolliet's earlier attempts at charting the Saint Lawrence River Valley were based on 46 voyages he made over a period of five years. This culminated in a large-scale regional chart, which was eventually compiled by Franquelin in 1685. The latter chart was, however, insufficiently accurate to safely navigate the entire river course.

## 5 SCIENCE AND POLITICS

The 1690s witnessed an increased military role in the North Atlantic, prompting calls for more accurate maps and sailing directions (Pritchard, 2004). The commonly used Dutch 'portolan'-style *pascaarten*, mostly dating from the early to mid-17th century, were not sufficiently reliable to sail up the Saint Lawrence River.[35] European navigators had developed portolan maps—from the Italian word *portolani*, a 'collection of sailing directions related to ports'—toward the end of the 13th century. They represented a novel type of marine charts based on direct observation using a mariner's compass (Lane, 1963). These maps, accurate even by modern standards, were used routinely until well into the 17th century (e.g., de Grijs, 2017).

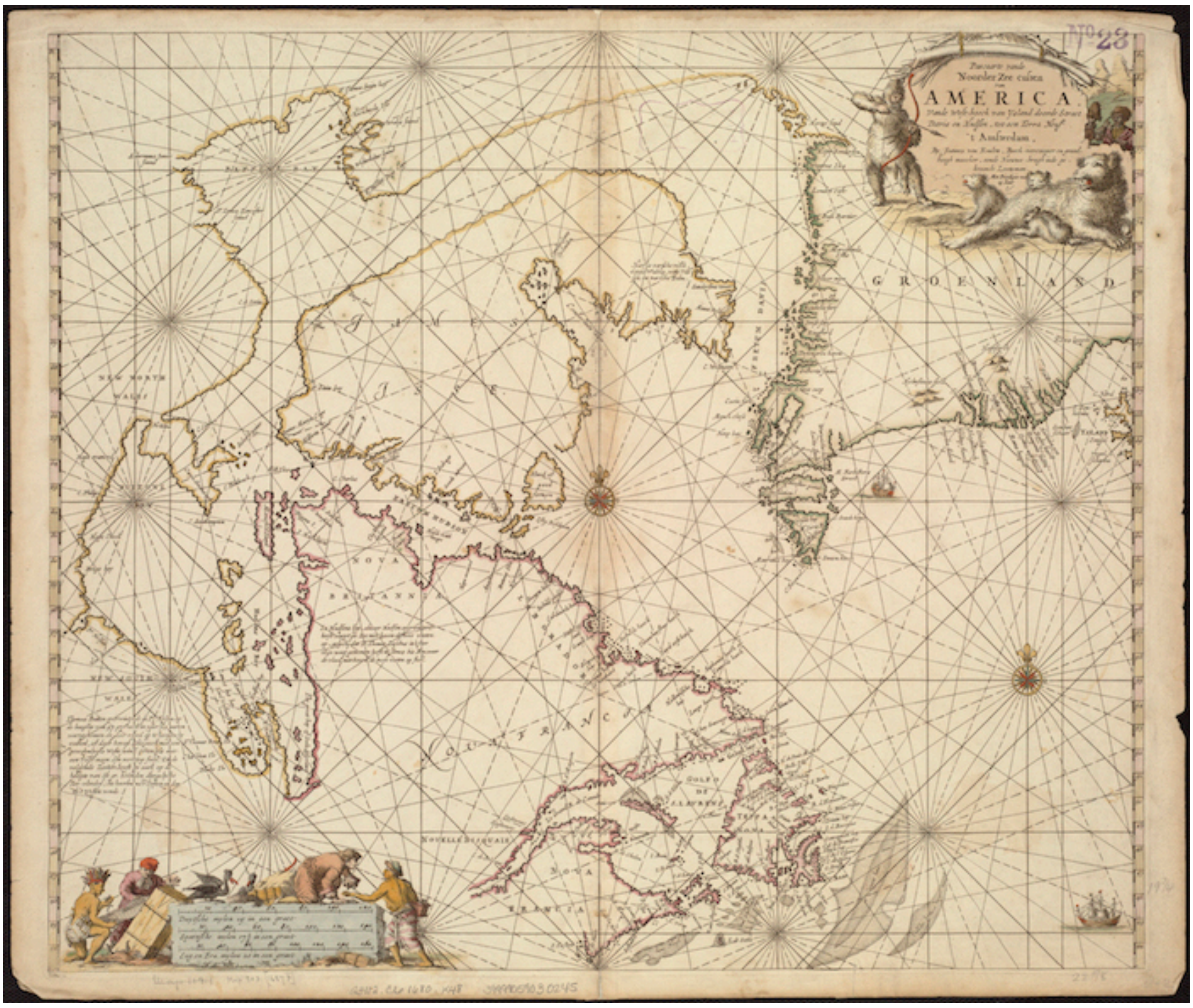

**Figure 6**: *Pascaarte vande Noorder Zee Custen van America, vande West-Hoeck van Ysland doorde Straet Davis en Hudson, tot aen Terra Neuf* ('Navigational chart of the North Atlantic Coast from Western Iceland through the Davis Strait and Hudson to Newfoundland'; in: van Keulen, 1697–1709; Norman B. Leventhal Map Center via Wikimedia Commons, CC-BY-2.0).

Figure 6 shows an excellent representation of one such *pascaarte*, authored by the Dutch cartographer Johannes van Keulen (1654–1715) between 1697 and 1709. It is based on an earlier navigation chart, the *Atlas van Loon* (1666), by Johannes Janssonius (1588–1664), combined with an expanded map from 1678 by the Amsterdam-based publisher and cartographer Hendrick Doncker (1626–1699). These charts were only meant to aid sailors in crossing the Atlantic and finding their bearings upon arrival near the opposite coast. Fairly rudimentary renderings of any inland waterways were based on previous English and French charts. They were insufficiently detailed for safe navigation. Despite attempts by individual navigators to collect directions while sailing across the North Atlantic, there was no single clearinghouse that could organize these efforts into a coordinated approach. This situation was not conducive to detecting problems and it actually encouraged personal rivalries to develop. In 1699, the first naval archives were established, leading to some coordination of the proliferation of maps and charts that had accumulated across government offices.

That same year, Deshayes sought permission to publish his 1686 chart of the Saint Lawrence River Valley. The French Secretary of State for the Navy, Jérôme Phélypeaux (1674–1747), Comte de Pontchartrain, forwarded Deshayes' request to the *Académie*.[36] In June 1699, the *Académie* formally endorsed the chart's publication and Deshayes obtained a *privilège* from the King to publish his chart.[37] It was published in 1702, although it may have been rendered on a reduced scale compared with the original (Litalien et al., 2007). Publication was undertaken "A Paris Chez N[icolas] de Fer dans l'Isle du Palais sur le Quay de l'Orloge à la Sphere Royale. Avec permission du Roi"—indeed, with the King's permission. The fact that a royal privilege was required upon endorsement by the *Académie* did not facilitate rapid dissemination of novel scientific results. In fact, it gave too much power and control to the institution for science to successfully contribute to seeking practical solutions to real-world problems—indeed to some extent contradicting the *Académie*'s *raison d'être*…

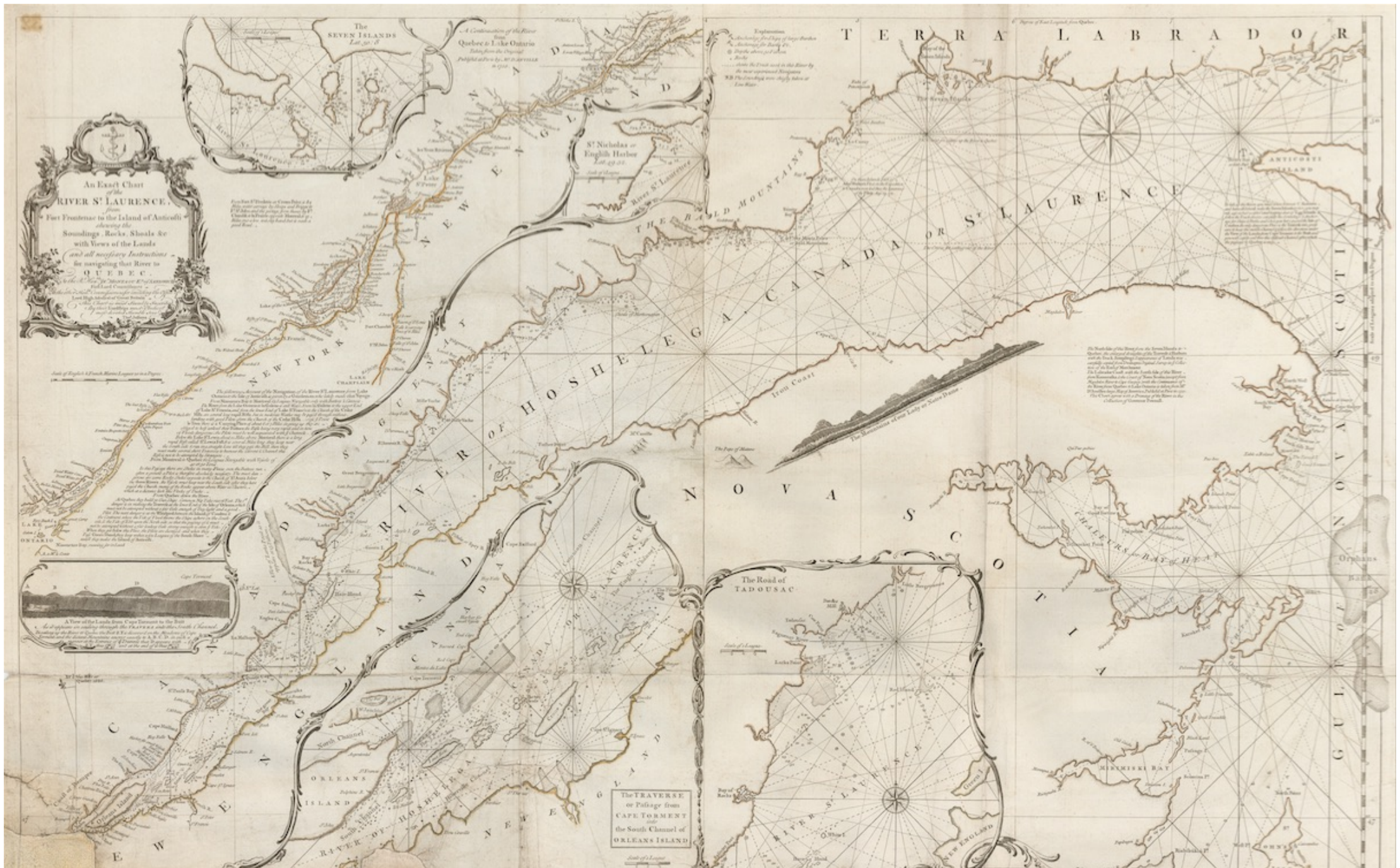

**Figure 7**: Thomas Jefferys' 1757 chart of the Saint Lawrence River Valley (reissued in 1775), based on Deshayes' 1715 precursor chart. (*Image provided by www.RareMaps.com, Barry Lawrence Ruderman Antique Maps, Inc.*; reproduced with permission)

## 6 LATER LIFE

Despite calls for new surveys[38] the War of the Spanish Succession (1701–1714) and associated military altercations with the English in North America,[39] as well as depleted navy finances, prevented this from happening. Deshayes' chart was reprinted after the death of King Louis XIV in 1715, but updated with sailing directions based on journal entries from the 1690s by captain Pierre le Moyne d'Iberville (1661–1706). It became a basic reference tool for the navigators and pilots of New France until the end of French rule in Québec in 1763, when France ceded control of its North American colonies to Great Britain and Spain. Subsequent charts of the Saint Lawrence River Valley—including those compiled by Richard Testu de la Richardière (1730–1741), Gabriel Pellegrin (1734–1755), Jean Baptiste Bourguignon d'Anville (1755), Jacques Nicolas Bellin (1757), Cook (1775) and others—used Deshayes' 1702/1715 chart as their baseline. Other cartographers also contributed independently around the same time. One particularly noteworthy example is the 1709 chart of Jean-Baptiste Decouagne (1687–1740), surveyor and military engineer, which was based on competent use of the latest techniques and instruments. In 1757, Deshayes' hydrographic map took on unexpected military and commercial importance. That year, the British firm of Thomas Jefferys copied it for use by British navigators (see Figure 7); its chart was subsequently reissued three times until 1775.

Meanwhile, under the guidance of Lieutenant Samuel J. Holland (1728–1801), Cook gained his first experience in hydrographic surveying and cartography, at Kennington Cove (Cape Breton Island, at the time known as Île Royale), using a plane-table to chart the Gulf of Saint Lawrence,[40] the Saint Lawrence River, and eventually also Newfoundland and Labrador (Skelton, 1954; Ritchie, 1978; Orchiston, 2017). Charting of the Saint Lawrence Seaway had become a high-priority military endeavour, given British attempts to capture Québec and the surrounding French territory. Their campaign was hampered by the treacherous waters of the Saint Lawrence River approach to Québec, for which Deshayes' updated 1715 chart was still the most detailed resource at the time.

Following Deshayes' sudden departure from Québec in November 1686, his name all but disappears from any formal correspondence until the beginning of the 18th century. It is thought that he returned to teaching mathematics in France until 1702, when he was dispatched once again across the Atlantic. The high quality of the chart resulting from his earlier period of residence in the colonies may well have contributed to his appointment as Royal Hydrographer of New France in 1702, as successor to both Franquelin and Jolliet. He arrived back in Québec later that year, tasked with teaching navigation and pilotage for the next four years at the Jesuits' Séminaire (Collège) de Québec as the colony's Professor of Hydrography. Louis Jolliet had died in 1700, leaving the education of navigation in the colonies in disarray. In 1703, Deshayes was promoted to the colony's deputy engineer. The following year, he proceeded to compile a new chart of the Saint Lawrence River's northern shore to visualize the expedition to Labrador undertaken by Augustin le Gardeur de Courtemanche (1663–1717), the Canadian soldier and ambassador.

In November 1707, Philippe de Rigaud (c. 1643–1725), Marquis de Vaudreuil, Governor-General of New France at the time, and one Mr. Begou, Intendant of La Rochelle, announced Deshayes' death in a joint letter to Minister Colbert. Although the parish records of Notre Dame de Québec do not contain the date of Deshayes' death, the royal notary Florent de la Cetière retained this detail in an inventory he compiled of Deshayes' personal effects. The cartographer died on 18 December 1706 in the Hôtel-Dieu (hospital) de Québec.

Deshayes left a modest inventory, including a surveyor's level with tripod, a quadrant, three compasses, some lenses, a ruler, and a geometric compass. He also left behind a 43-volume library that stood out among others in the colony because it did not include religious books, while the scientific manuscripts were of very high quality—including five books about astronomy,[41] a number of treatises about tides and meteorology, and seven books about navigation and pilotage, including a copy of the *Théorie des manœuvres des vaissaux*. He also owned some copies of the *Connoissance des temps*, an almanac established in 1679 and subsequently published by the *Académie* since 1702. This publication is noteworthy, because it was entirely based on exact astronomical information and measurements, without any reliance on the astrological notions still so commonly invoked at the time.

## 7 A SIGN OF THE TIMES

Throughout Deshayes' active career, Colbert became one of the most important drivers of innovation in mapmaking. He stood at the basis of the foundation of the *Académie Royale des Sciences*. He also contributed significantly to what later became known as the *Dépôt des cartes et plans de la Marine*. Although the institution was formally established only in 1720, Colbert's hand in its creation is evident from around 1682, when he appointed a technician tasked with compiling and organizing cartographic documentation received by his Ministry.

The world was about to change irrevocably. Trade routes expanded significantly, in turn necessitating more accurate maps (for a comprehensive review of these developments, see de Grijs, 2017).[42] Aristotelian belief systems gave way to approaches we now collectively refer to as the scientific method, although the Jesuit missionaries retained their traditional cosmology for a while. Indeed, the 17th century Scientific Revolution had already managed to gain a robust foothold in the nascent community of scholars and natural philosophers, with more to come in the 18th century, which had only just begun. Most importantly in the context of our current narrative, the size and shape of the Earth became ever more tangible. By the end of the century, the English clockmaker John Harrison (1693–1776) solved the intractable 'Longitude Problem' through innovative design and advances in metallurgy. Positions could finally be determined reliably.

## 8 NOTES

[1] On 3 July 1608, Samuel de Champlain (1574–1635), the French explorer and diplomat now known as the Father of New France, had established the town of Québec. The entire population of New France consisted of around 11,000 residents in 1685. Population growth in the 17th and 18th centuries proceeded slowly, with 18,500 residents recorded in 1713 and 55,000 in 1754. In that latter year, the population of Québec City had only just reached 8000 (Kerr, 1966).

[2] Hydrographers measure and describe physical features of bodies of water and the land–water interface, primarily for navigational purposes.

[3] The territory of New France consisted of five colonies, including Canada, Hudson's Bay, Acadia, Plaisance (on Newfoundland) and Louisiana. Canada, in turn, was subdivided into the districts of Québec, Trois-Rivières and Montréal.

[4] Deshayes was not the first scholar to attempt longitude determination in Canada (Jarrell, 1988), but his results were the most accurate yet. Prior to Deshayes' determination, Captain Thomas James (1593–1635) had observed the 29 October 1631 lunar-eclipse timings, which the mathematician Henry Gellibrand (1597–1637) attempted to use, in combination with his own observations in London, to achieve the same for the longitude of Charlton Island. He also used stellar occultations (where the Moon eclipsed a star in the constellation Corona Borealis) to obtain an independent estimate. Although his results were very close to the present-day value, Jarrell (1988) suggests that this may be owing to a combination of errors cancelling each other. Around the same time, the Jesuit priest Pierre Le Jeune (1591–1664) attempted a similar feat, but with a rather poor accuracy of 19°.

[5] Navigation of the river and exploration of new lands would drive the economy under the French administration, which lasted until the Battle of the Plains of Abraham, just outside the walls of Québec, on 13 September 1759, between Generals Montcalm and Wolfe.

[6] On 17 June 1685, Colbert provided detailed instructions to the intendant and interim governor of New France, Jacques de Meulles (d. 1703), as regards the work that was to be undertaken by Deshayes. He also requested that his envoy be provided with a vessel to travel down the river (Roy, 1916).

[7] Although several sources emphasize Deshayes' orders to obtain astronomical observations (e.g., Roy, 1895; Roy, 1916), the type of observations he was meant to obtain is not specified. It is likely that he was tasked with determining latitudes of locales along the Saint Lawrence River using astronomical observations.

[8] Unless otherwise indicated, all translations from French are my own.

[9] Although determination of one's latitude was much easier than obtaining longitude, on land solar altitude measurements were affected by a combination of factors that rendered them uncertain, including the difficulty of determining the exact time of local noon, instrumental and observer errors, and not having corrected for atmospheric or instrumental refraction.

[10] Although the much more accurate sextant was not used in practice until around 1731, by John Hadley (1682–1744) and Thomas Godfrey (1704–1749), in 1685 one could obtain latitude determinations to

accuracies of several seconds of arc using multiple measurements from different base stations. Deshayes was unable to do so because of time pressure.

[11] Although maintaining the accuracy of a clock *at sea* remained a significant challenge until John Harrison's construction of his longitude clock 'H4,' accurate timekeeping *on land* had been achieved well before Deshayes' observation of the December 1685 lunar eclipse (for a recent exposition, see de Grijs, 2017). For instance, in April 1642, the Italian astronomer and Jesuit priest Giovanni Battista Riccioli (1598–1671) and his associates kept a "seconds pendulum going for 24 hours, counting 87,998 oscillations" (*Almagestum Novum*, 1651), close to the 86,400 oscillations expected. In the following months, he repeated this experiment twice and managed to improve the accuracy to better than 0.69% (Heilbron, 1999; Meli, 2006). Christiaan Huygens' invention of the pendulum clock in 1656 was aimed at automating this process by adapting the existing clockwork mechanism to count the pendulum swings and to sustain its motion in the presence of friction.

[12] It is likely that Deshayes used Cassini's new lunar map of 1679 for his precise lunar eclipse observations, although there are no records to support this suggestion. Nevertheless, the French Jesuit missionary delegation to Siam (present-day Thailand) of 1685, which also observed the 10–11 December 1685 lunar eclipse (see note 15), is known to have used Cassini's map for their lunar-eclipse timings (Gislén et al., 2018). Therefore, it seems logical that Deshayes also had access to a copy of the same map, given that he had set sail several years after the map's completion.

[13] https://en.wikipedia.org/wiki/List_of_17th-century_lunar_eclipses; https://eclipse.gsfc.nasa.gov/LEcat5/LE1601-1700.html [accessed 29 October 2018].

[14] 'Absolute' time measurements, used for calibration of the clocks used at either location, could be achieved by reference to the well-known and easily measurable ephemerides of Jupiter's moons and the daily motion of the Sun across the sky.

[15] A delegation of French Jesuit missionaries, including the astronomer Guy Tachard (1651–1712), observed the 10–11 December 1685 lunar eclipse from Lop Buri, Siam (Tachard, 1686; Orchiston et al., 2016). They used their lunar-eclipse timing observations to determine the longitude of their observing site as 121°02′ East of the Canary Island of Ferro (El Hierro), a commonly used reference meridian at the time (e.g., de Grijs, 2017). The modern longitude determination of Lop Buri is 118°42′ East of El Hierro.

[16] This document is held in the archives of the Musée de Civilisation in Québec City; *Polygraphie*, 2, No. 34; see also Trudel (2008).

[17] It remained an uncontested reference for half a century, until Michel Chartier de Lotbinière (1723–1798), King's Engineer in the Colonial Regular Army, obtained an improved measurement to within 56' of the real distance. This new precision was proudly announced to the *Académie* by Roland-Michel Barrin (1693–1756), Marquis de La Galissonière, a keen supporter of research into geography and the natural sciences: "I placed Quebek [sic] at 46 degrees 48 minutes of latitude, & at 72 degrees 38 minutes west longitude of the meridian of Paris, following the observations made there in 1754, by Mr. de Lo[t]biniere, of the troops of Canada, well versed in astronomy" (Bellin, 1755).

[18] https://tools.wmflabs.org/geohack/geohack.php?pagename=City_Hall_of_Quebec_City&params=46.81391_N_71.207954_W_ [accessed 29 October 2018].

[19] Reference number 51718 of the *Commission de toponymie du Québec*; http://www.toponymie.gouv.qc.ca/ct/ToposWeb/fiche.aspx?no_seq=51718 [accessed 29 October 2018].

[20] *Géoportail: le portail national de la connaissance du territoire mis en oeuvre par l'IGN*; https://www.geoportail.gouv.fr/accueil?c=2.3508,48.8567&z=0.000316906&l=GEOGRAPHICALGRIDSYSTEMS.MAPS.3D$GEOPORTAIL:OGC:WMTS@aggregate(1)&permalink=yes [accessed 29 October 2018].

[21] Deshayes also produced a *Carte des côtes habitées du Canada par paroisses et par seigneuries*, that is, a detailed cadastral map of the Canadian colony, which included the names of all *seigneuries* granted by 1686. This latter map was reproduced in 1893 by E. Dufosse in Paris in a compendium of North American maps, the *Recueil de cartes, plans et vues relatifs aux États-Unis et au Canada, 1651–1731*.

[22] The Deshayes referred to in the 1660s may or may not have been the same person as the hydrographer responsible for charting the Saint Lawrence River (Dew, 2010; footnote 17). Jarrell (1988) suggests that Deshayes may have been one of Cassini's students at the Observatoire de Paris, although he does not provide any evidence to substantiate this claim. On the other hand, Danson (2006) suggests that this may have been a certain Louis des Hayes instead, although without providing any provenance for that claim.

[23] Deshayes took both his own, newly invented 'instrument' (about which little is known) and two clocks, most likely provided by Huygens, on the voyage to Acadia (de Grijs, 2017).

[24] Deshayes adopted the royal *toise* as his benchmark, which equaled 6 *pieds*; a *pied* (making up 12 *pouces*) was equal to 12.789 English inches (Daumas, 1953; see also Pritchard, 1979).

[25] In Pritchard (1979), the editor has added a note suggesting that this 'bark' may have been a schooner rigged two-masted vessel.

[26] As the British Admiralty's fleet did, led by Lord Colville, when it sailed through the Narrows to attack Québec 73 years later.

[27] Deshayes relied to some extent on sounding measurements by his crew, but the latter were less than reliable. On his final chart, he indicated those measurements using Roman numerals to distinguish them from his own (Pritchard, 1979).
[28] The Paris-based Englishman Michael Butterfield was Louis XIV's designer and mathematical instrument maker. He was known as the finest craftsman of his day (Daumas, 1953).
[29] Such a sighting device or pointer for determining directions or measuring angles is known as an 'alidade.'
[30] The folding edge rule was graduated into degrees, which clamped down the sheet of paper onto the plane-table surface.
[31] *Carte marine de l'embouchure de la rivière de S. Laurens levée de cap en cap jusqu'a Québec verifiée par plusieurs Observations. Plus le cours de cette rivière au-dessus de Québec jusqu'au Lac Ontario. Par le sieur Des Hayes Hydrographe*.
[32] *Histoire de l'Académie royale des sciences, 1699* (Paris, 1702), p. 86.
[33] *Carte marine de la Rivière de Québec par le Sr. Deshayes, 1686. Ou recueil de ce qui sert à la navigation particulière de cette rivière et de ce qui peut contribuer à la metode* [sic] *générale de lever et dresser les cartes marines.*
[34] http://www.oldworldauctions.com/archives/detail/92-099.htm [accessed 20 June 2018].
[35] Note that Pritchard adopts the unusual spelling 'paescart' instead of the more common 'pascaarte'.
[36] *Archives de la Marine*, Ser. $B^2$, 140, p. 111: Pontchartrain, Jérôme – Abbé Bignon; April 1699.
[37] The Académie's certificate is included in its annual report, *Histoire de l'Académie royale des sciences, 1699* (Paris, 1702), p. 86; it was reprinted in 1932 in the *Bulletin des recherches historiques*, XXXVIII, p. 281. Deshayes' permission is found in the *Archives de la Marine*, Ser. $B^2$, 140, p. 329: Pontchartrain, Jérôme – Deshayes, Jean; 3 June 1699.
[38] In 1702 one captain complained, "once more, my Lord, the charts are worthless" (Pritchard, 2004: 15; transl.) following his unsuccessful attempt at locating Cape Sable at the southwestern extremity of Nova Scotia.
[39] In 1713, King Louis XIV was forced to cede Acadia to England, for instance.
[40] Although Cook's main chart of the Saint Lawrence River was not published until 1775, an early chart drawn by Cook, "*A Plan of the Traverse or Passage from Cape Torment into the South Channel of Orleans*", was completed in 1759–1760 (British Library, Add. MS 31360, f. 14).
[41] Deshayes was not the first astronomer in France's North American colonies (see Jarrell, 1988); as early as 30 November 1618, Joseph Le Caron—missionary to the Huron nation in Georgian Bay, on the Ottawa River—observed a comet from Montagnet, near Tadoussac, where he was due to spend the winter (Sagard, 1636). This may have been one of the two comets Johannes Kepler observed on 29 November 1618 (Broughton, 1976).

## 9 ACKNOWLEDGEMENTS

I thank Noriyuki Matsunaga for his great hospitality during a week-long visit to the University of Tokyo, where most of this article was composed. The suggestions received from the reviewers enabled me to place my narrative in a more complete, global context.

## 10 REFERENCES

Archibald, T., Davison, B., and Lysne, M., 2009. *Mathematics in Canada to 1980: An historical assessment*, Canada Museum of Science and Technology; https://documents.techno-science.ca/documents/HistoricalAssessment-HA-MathematicsinCanadato1980s-2009.pdf [accessed 20 June 2018].

Bellin, J.N., 1755. *Carte de l'Amérique septentrionale depuis le 28 degré de latitude jusqu'au 72*. Paris: Imprimerie de Didot.

Broughton, P., 1976. Canadian Comet Discoveries, *J. R. Astron. Soc. Can.*, 70, 311–319.

Chartrand, L., Duchesne, R., and Gingras, Y., 1987. *Histoire de sciences au Québec*, Montréal: Boréal.

Danson, E., 2006, *Weighing the World: The Quest to Measure the Earth*. Oxford: Oxford Univ. Press, p. 27.

Daumas, M., 1953. *Les instruments scientifiques aux XVII^e^ et XVIII^e^ siècles*, Paris: P.U.F.

de Grijs, R., 2017. *Time and Time Again: Determination of longitude at sea in the 17th century*. Bristol, UK: IOP Publishing.

Delanglez, J., 1943. Franquelin, mapmaker, *Mid-America*, 25, 36–37.

Deshayes, J., 1686. *Carte marine de la Rivière de Québec par le Sr. Deshayes, 1686. Ou recueil de ce qui sert à la navigation particulière de cette rivière et de ce qui peut contribuer à la metode* [sic] *générale de lever et dresser les cartes marines*, f. 17; transl.:

Pritchard (1979: 131). See also http://hdsquebec.org/cartes-maritimes.htm [accessed 20 June 2018].
Dew, N., 2010. Scientific travel in the Atlantic world: the French expedition to Gorée and the Antilles, 1681–1683, *Brit. J. Hist. Sci.*, 43, 1–17.
Gagnon, E., 1900–1901. Louis Jolliet, *Rev. Canad.*, 37, 464–465.
Gauja, P., 1949. L'Académie Royale des Sciences (1666–1793), in: *Revue d'histoire des sciences et de leurs applications*, 2, 293–310.
Gislén, L., Launay, F., Orchiston, W., Orchiston, D.L., Débarbat, S., Husson, M., George, M., and Soonthornthum, B., 2018. Cassini's 1679 map of the Moon and French Jesuit observations of the lunar eclipse of 11 December 1685. *Journal of Astronomical History and Heritage*, 21, 211–225.
Gosselin, A., 1897. Encore le P. de Bonnécamps 1707–1790, *Trans. R. Soc. Can.*, 2nd ser., XII, sect. 1, p. 98.
Hamilton, A.C., and Sebert, L.M., 1996. *Significant Dates in Canadian Surveying Mapping and Charting*, Ottawa: Geomatica Press.
Harper, W.L., 2011. *Isaac Newton's Scientific Method: Turning Data into Evidence about Gravity & Cosmology.* Oxford: Oxford Univ. Press.
Heidenreich, C.E., 2006. La cartographie jusqu'en 1763, in: *Histoire de la cartographie au Canada*; https://www.encyclopediecanadienne.ca/fr/article/cartographie-histoire-de-la/ [accessed 18 June 2018].
Heilbron, J.L., 1999. *The Sun in the Church: Cathedrals as Solar Observatories.* Cambridge, MA: Harvard Univ. Press.
Henrion, D., and Deshayes, J., 1681. *L'usage du compas de proportion*, Paris: Chez l'Autheur, au bout du Pont-Neuf, proche le Bureau du Grenier à Sel. et chez R. J. B. de la Caille, rue S. Jacques, aux trois Cailles: https://archive.org/details/bub_gb_HcNyW9KgkDsC/page/n5.
Jarrell, R.A., 1988. *The Cold Light of Dawn. A History of Canadian Astronomy*, Toronto: Univ. of Toronto Press.
Jones, P.S., 1967. The History of Mathematical Education, *Am. Math. Monthly*, 74, 38–55.
Kerr, D.G.G., 1966. *A Historical Atlas of Canada*, 2nd ed. Toronto: Don Mills, Thomas Nelson.
Lane, F.C., 1963. The Economic Meaning of the Invention of the Compass, *Am. Hist. Rev.*, 68, 615ff.
Leybourn, W., 1722. *The Compleat Surveyor: Or, The Whole Art of Surveying of Land, By a New Instrument Lately Invented; as Also by the Plain Table, Circumferentor, the Theodolite as Now Improv'd, Or by the Chain Only*, London: Printed for Samuel Ballard at the Blue Ball, and Aaron Ward at the King's Arms in Little Britain, and Tho. Woodward at the Half-Moon against St. Dunstan's Church in Fleet Street.
Litalien, R., Palomino, J.-F., and Vaugeois, D., 2007. *La Mesure d'un Continent: Atlas historique de l'Amérique du Nord 1492–1814*, Paris: Presses de l'Université Paris-Sorbonnes.
Meli, D.B., 2006. *Thinking with Objects: The Transformation of Mechanics in the Seventeenth Century*. Baltimore, MD: Johns Hopkins Univ. Press.
O'Connor, J.J., and Robertson, E.F., 2012. Jean Richer, *MacTutor History of Mathematics Archive*; http://www-history.mcs.st-and.ac.uk/Biographies/Richer.html [accessed 20 June 2018].
Olmsted, J.W., 1960. The Voyage of Jean Richer to Acadia in 1670: A Study in the Relations of Science and Navigation under Colbert, *Proc. Am. Philos. Soc.*, 104, 612–634.
Orchiston, W., 2017, Cook, Green, Maskelyne and the 1769 transit of Venus: The legacy of the Tahitian observations. *Journal of Astronomical History and Heritage*, 20, 35–68.
Orchiston, W., Orchiston, D.L., George, M., and Soonthornthum, B., 2016. Exploring the first scientific observations of lunar eclipses made in Siam. *Journal of Astronomical History and Heritage*, 19, 25–45.
Palomino, J.-F., 2009 Entre la recherche du vrai et l'amour de la patrie: cartographier la Nouvelle-France au XVIII$^{e}$ siècle, *Revue de Bibliothèque et Archives nationales du Québec: Cartographie*, 1, 84–99.
Pritchard, J.S., 1969 (revised 1982). Deshayes, Jean, *Dictionary of Canadian Biography*, 2, Univ. Toronto/Univ. Laval; http://www.biographi.ca/en/bio/deshayes_jean_2E.html [accessed 20 June 2018].
Pritchard, J.S., 1979. Early French Hydrographic Surveys in the Saint Lawrence River, *Int'l*

*Hydrogr. Rev*., LVI, 125–142.
Pritchard, J.S., 2004. Hydrography in New France, in: *Charting Northern Waters: Essays for the Centenary of the Canadian Hydrographic Service*, Glover, W. (ed), Montreal and Kingston: McGill–Queens Univ. Press, 10–21.
Ritchie, G.S., 1978. Captain Cook's influence on hydrographic surveying. *Pacific Studies*, 1, 78–95.
Roy, J.E., 1895. La Cartographie et l'Arpentage sous le Régime Français, *Bull. Res. Hist*., I, 33–40.
Roy, P.-G., 1916. Jean Deshayes, hydrographe du Roi, *Bull. Res. Hist*., XXII, 129–138.
Sagard, G., 1636. *L'Histoire du Canada et voyages que les Frères Mineurs Recollects y ont faicts pour la conversion des Infidelles*, A Paris, Chez Claude Sonnius, rue S. Jacques, à l'Escu de Basle, & au Compas d'or.
Saunders, E.S., 1984. Louis XIV: Patron of Science and Technology, Libraries Research Publ., No. 46; http://docs.lib.purdue.edu/lib_research/46 [accessed 13 November 2018].
Skelton, R.A., 1954. Captain James Cook as a hydro-grapher. *Mariner's Mirror*, 40, 92–119.
Sofroniou, A., 2016. European Influences, in: *Triangle of Education Training Experience*. Lulu Press.
Tachard, G., 1686. *Voyage de Siam des Pères Jésuites Envoyés par le Roi aux Indes & à la Chine.* Paris, Seneuze et Horthemels.
*The Quarto*, June 1974, No. 105, p. 3. Ann Arbor, MI: Clements Libr., Univ. of Michigan; http://clements.umich.edu/Quarto/Quarto_1st%20series_105,%20June%201974.pdf [accessed 29 October 2018].
Trudel, C., 2008. Trouvailles: Un fleuve périlleux, *Exploration de Bibliothèque et Archives nationales du Québec (BAnQ)*. http://cltr.blogspot.com/2008/04/un-fleuve-prilleux.html [accessed 18 June 2018].
van Keulen, J., 1697–1709. *De Nieuwe Groote Lichtende Zee-Fackel, Behelsende 't Eerste, 't Tweede, 't Darde, 't Vierde, 't Vijfde of 't Laetste Deel … beschruvinge, van alle bekende Haavens … door J. van Loon, en C. J. Vooght*, Amsterdam.
Varin, Deshayes, J., and de Glos, G., 1682. *Observations astronomiques faites au Cap Verd, en Afrique, et aux Isles de l'Amerique*; reprinted in *Mémoires de l'Académie Royale des Sciences Depuis 1666 jusqu'a 1699* (Paris, 1729), VII, 431–459.
Vogel, C., 2010. The Suppression of the Society of Jesus, 1758–1773, in: *European History Online* (EGO), published by the Institute of European History (IEG), Mainz, Germany: http://www.ieg-ego.eu/vogelc-2010-en [accessed 9 November 2018].